\documentclass[12pt]{article}
\usepackage[T2A]{fontenc}
\usepackage[utf8]{inputenc}
\usepackage[english]{babel}
\usepackage{amsmath}
\usepackage{amssymb}
\usepackage{amsthm}
\usepackage[affil-it]{authblk}
\usepackage{upgreek}
\usepackage{hyperref}
\hypersetup{
	colorlinks=true,
	linkcolor=red,
	citecolor=blue,
	urlcolor=blue
}

\usepackage[a4paper, total={6in, 9.6in}]{geometry}
\newcommand{\der}[2]{\dfrac{\mathrm{d} #1}{\mathrm{d} #2}}

\def\L{\mathcal{L}}     
\def\O{{\mathcal O}}
\def\Tr{{\rm Tr}}       

\newcommand{\bk}{{\bf k}}
\newcommand{\D}{{\mathcal D}}

\newtheorem{theorem}{Theorem}%
\newtheorem{remark}{Remark}%

\title{\textbf{Quantum control by the environment: Turing uncomputability, Optimization over Stiefel manifolds, Reachable sets, and Incoherent GRAPE}}

\date{}

\author[1,2,**]{Alexander Pechen}

\affil[1]{\it \normalsize Department of Mathematical Methods for Quantum Technologies,\par
	Steklov Mathematical Institute of Russian Academy of Sciences,\par
	8~Gubkina str., Moscow, 119991, Russia, }
 \affil[2]{\it 
	University of Science and Technology MISIS,\par
	6~Leninskiy prosp., Moscow, 119991, Russia;}
\affil[**]{apechen@gmail.com, \href{http://www.mathnet.ru/eng/person17991}{mathnet.ru/eng/person17991}}

\begin{document}
\maketitle
	
\begin{abstract} 
The ability to control quantum systems is necessary for many applications of quantum technologies ranging from gate generation in quantum computation to NMR and laser control of chemical reactions. In many practical situations, the controlled quantum systems are open, i.e., interacting with the environment. While often influence of the environment is considered as an obstacle for controlling the systems, in some cases it can be exploited as a useful resource. In this note, we briefly review some results on control of open quantum systems using environment as a resource, including control by engineered environments and by non-selective measurements, Turing uncomputability of discrete quantum control, parametrization of Kraus maps by points of the Stiefel manifolds and corresponding Riemanninan optimization, control by dissipation and time-dependent decoherence rates, reachable sets, and incoherent GRAPE (Gradient Ascent Pulse Engineering) --- inGRAPE --- for gradient-based optimization.\footnote{This work is based on the talk presented at the 15th International Conference “Micro- and Nanoelectronics --- 2023”  October 2--6, 2023, Zvenigorod, Russia, and prepared for the Proceeding of the Conference. Since it is a brief overview of several concrete results, we do not provide here a detailed discussion of various related approaches and other works. A more detailed overview of other related approaches can be found in the cited works.}
\end{abstract}
	
\noindent{{\bf Keywords:} quantum control, control by dissipation, open quantum systems, Riemannian optimization, time-dependent decoherence rates
	
\section{Introduction}\label{sec1}

Quantum control plays am important role in laser chemistry, optical control of molecular processes~\cite{RiceZhaoBook,TannorBook,Shapiro_Brumer_2011} and modern quantum technologies~\cite{Schleich2016,Acin2018,Koch2022,Mancini_Manko_Wiseman_2005,Gough_2012}. Examples of its application range from robust or time-optimal generation of quantum gates in the presence of dissipation to steering a spin 1/2 particle to a maximally mixed state for NMR, laser control of chemical reactions, etc. In many experimental situations, the controlled quantum system is an open quantum system. i.e. interacting with the environment. While often the environment acts as an obstacle for controlling the systems, in some cases it can be exploited as a useful resource. Here we review the approach for such use based on the works~\cite{PechenRabitzPRA2006,Pechen2011}, and discuss various related results including:
\begin{itemize}
    \item Control by engineered environment and by non-selective quantum measurements;
	\item Turing uncomputability of discretized quantum control;
	\item Riemannian optimization over complex Stiefel manifolds for control of open quantum systems;
	\item Reachable sets for an open qubit;
	\item Incoherent GRAPE (inGRAPE) optimization.
\end{itemize}

\section{Environment as a resource, incoherent control and time-dependent decoherence rates}\label{secIncoherent}

There are many works and approaches which consider the environment as a useful resource. Here we outline the approach to  engineered environment based on incoherent control which was developed in 2006~\cite{PechenRabitzPRA2006} and studied in~\cite{Pechen2011}. In this approach, state of the environment is used as a control. A similar idea of using non-selective quantum measurements as controls was proposed in 2006~\cite{PechenIlinShuangRabitzPRA2006}, where an exact analytical solution for the two-level case was also obtained.  Note that non-selective measurements and the environment are very close (and basically the same) mechanisms.  Usually in theory of open quantum systems state of the environment is considered as thermal (Gibbs) state. However, this assumption is too restrictive as the environment can be prepared in more general non-thermal, non-equilibrium and even time-dependent states. State of the environment obviously affects the dynamics of a quantum system immersed in this environment and hence can be used as a control.

In many cases, state of the environment is a Gaussian state. It means that multiparticle correlation functions of particles of the environment are expressed in terms of the first order (mean) and second order moments. A Gaussian mean zero state of an environment can be characterized by the so called spectral density of the environment, which determines the two-point correlation function of its particles. If particles of the environment are characterized only by momentum $\bk$ and the environment is described by creation and annihilation operators $a^\pm(\bk)$, then its two-point correlation function is
\[
\langle a^+(\bk)a^-(\bk')\rangle=n(\bk)\delta(\bk-\bk')
\]
where $n(\bk)\ge 0$ is spectral density (density of particles of the environment with momentum $\bk$) and $\delta(\bk-\bk')$ is the Dirac delta-function. In general, spectral density can be any time-dependent non-negative  function $n(\bk,t)\ge 0$.

An example is the environment formed by incoherent photons. A time-evolving distribution of photons induces generally time-dependent decoherence rates of the system which is immersed in this photonic environment, so that under certain approximations master equation for the system density matrix $\rho(t)$ can be considered as
\begin{equation}\label{Eq:ME2}
	\der{\rho(t)}{t} = \L^{u, n}_t\rho(t):= -i [H^{u, n}_t, \rho(t)] + \varepsilon\underbrace{\sum_k\gamma_k(t) \D_k\rho(t)}_{\displaystyle\D^n_t\rho(t)},\quad \rho(0) = \rho_0.
\end{equation}
Without controls and for thermal environment such master equation was derived, e.g. in the weak coupling limit~\cite{Davies1976,Accardi_Volovich_Lu}. Here $H^{u, n}_t$ is the Hamiltonian generally depending on coherent $u$ and incoherent $n$ controls, $\varepsilon$ is the strength of interaction of the system with the environment,  $k=(i,j)$ is the multiindex, $\gamma_k(t)=\gamma_{ij}(t)$ are generally time-dependent decoherence rates, $\D_k=\D_{ij}$ is the dissipator for transition between states $|i\rangle$ and $|j\rangle$. The rate of decoherence for the transition between system states $|i\rangle$ and $|j\rangle$ with transition frequency $\omega_{ij}=E_{j}-E_{i}$ (here $E_{i}$ is the energy of the system state $|i\rangle$) is density- (and hence generally time-) dependent~\cite{PechenRabitzPRA2006},
\begin{equation*}
	\gamma_k(t)=\gamma_{ij}(t) = \pi\int \mathrm{d} {\bf k} \,\delta(\omega_{ij}-\omega_{\bf k})|g({\bf k})|^2(n(\bf k,t)+\kappa_{ij}),\quad i,j = 1, \ldots, N.
\end{equation*}
Here $\kappa_{ij}=1$ for $i>j$ and $\kappa_{ij}=0$ otherwise, $\omega_{\bf k}$ is the dispersion law for the bath (e.g., $\omega=|{\bf k}|c$ for photons, where ${\bf k}$ is photon momentum and $c$ is the speed of light), and $g({\bf k})$ describes coupling of the system to $\bf k$-th mode of the photonic reservoir. For $i>j$, the summand $\kappa_{ij}=1$ describes spontaneous emission and $\gamma_{ij}$ determines rate of both spontaneous and induced emission between levels $i$ and $j$. For $i<j$, $\gamma_{ij}$ determines rate of induced absorption.

The idea of incoherent control is to use functions $n_{\omega_{ij}}(t)\ge 0$ as controls to manipulate the system. To answer how rich can be such control, it was shown in~\cite{Pechen2011} that such control can approximately generate, together with ultrafast coherent control, arbitrary density matrices of generic $N$-level open quantum systems, and hence can approximately realize complete state controllability in the set of all density matrices --- the strongest degree of quantum state control. 

Note that such control can induce in a controllable way various non-unitary dynamics of open quantum systems. Recall that most general dynamics of quantum systems are described by completely positive trace preserving maps (CPTP) called as Kraus maps or quantum channels\footnote{Sometimes not-completely positive dynamics is also considered~\cite{Pechukas_1994,Shaji_Sudarshan_2005}, but here we do not consider this case.}. A CPTP map is a map which is
\begin{itemize}
	\item Linear: $\Phi(\alpha\rho +\beta\sigma)=\alpha\Phi(\rho)+\beta\Phi(\sigma)$ for any $\alpha,\beta\in\mathbb C$.
	\item Trace preserving: $\Tr\Phi(\rho)=\Tr\rho$.
	\item Completely positive: for any $l\in N$: the map $\Phi\otimes{\mathbb E_l}\ge 0$, where $\mathbb E_l$ is the identity map in $\mathbb C^{l\times l}$.
\end{itemize}

Any CPTP map of an $N$-level quantum system has (a non-unique) Kraus operator-sum representation (OSR)
\begin{align}
	\Phi(\rho)=\sum\limits_{i=1}^{N^2} K_i\rho K^\dagger_i, \label{Kraus1}
\end{align}
where Kraus operators satisfy the consraint
\begin{align}
	\sum\limits_{i=1}^{N^2} K^\dagger_i K_i=\mathbb I_N\label{Kraus2}
\end{align}
where $\mathbb I_N$ is $N\times N$ identity matrix.

\section{Uncomputability of discretized quantum control}\label{secUncomputability}

While a combination of coherent and incoherent controls allows to approximately produce complete density matrix controllability (including arbitrary mixed states) of open quantum systems~\cite{Pechen2011} if we have unlimited controls, a natural question is about general (Turing) computability of quantum control tasks provided we have a finite number of controls.  Turing uncomputability (undecidability) was found for various important in quantum physics problems~\cite{Noce_Romano_2022}. Here we discuss the result of~\cite{Bondar_Pechen_2020} on uncomputability of the problem whether a finite sequence of Kraus maps from a given set exists which steers an initial quantum state to a target quantum state. We do not give here detailed references to other results on uncomputability topics for quantum systems, some of them can be found in~\cite{Bondar_Pechen_2020}.

To formulate our problem note that any coherent $u$ and incoherent $n$ controls applied to a quantum system during some time $T$ induce transformation of its states which is represented by some Kraus (CPTP) map $\Phi_{u,n,T}$. Suppose we have a finite number $K$ of various elementary controls $(u_1,n_1),\dots, (u_K,n_K)$ applied to the system. They induce some Kraus map transformations $\Phi_1,\dots,\Phi_K$. Suppose we can combine these controls in a sequence  in any order, including repeating them if necessary any finite number of times. In addition, we assume that all relevant physical quantities are described by rational numbers. 
\begin{remark} Both assumptions --- use of finite number of controls (e.g., in simplest case 'ON' and 'OFF') and description of the physical quantities and experimental results by rational numbers are exactly that occur in experiments.
\end{remark}
		
Consider the following question: Is there for given initial and target density matrices $\rho_{\rm i}$ and $\rho_{\rm f}$ a sequence $i_1,\dots, i_M$ (which can include repetitions) of finite length such that
\[
\Phi_{i_M}\circ\dots\circ\Phi_{i_1}(\rho_{\rm i})=\rho_{\rm f}?
\]
Any such question has either positive or negative answer. But is there a single algorithm which gives this answer for an input consisting of system, set of elementary controls, initial and target states?

As was shown in~\cite{Bondar_Pechen_2020}, there  is no algorithm (Turing machine) which answers this question for all such problems. The proof was given via connection with Diophantine equations and Hilbert tenth problem~\cite{MatiyasevichBook}.

\begin{remark} The negative result  of~\cite{Bondar_Pechen_2020} is also applied to the case of only unitary maps, i.e., if each elementary Kraus map $\Phi_k$, $k=1,\dots,K$, is a unitary transformation. 
\end{remark}
The negative result  implies that there is no single algorithm for all the problems of this type. But it does not exclude that for some subclasses of problems various algorithms do exist.

\section{Riemannian optimization over Stiefel manifolds}\label{secStiefel}

Optimization theory based on Newton and conjugate gradient algorithms on the real Grassmann and Stiefel manifolds was developed in the fundamental work~\cite{Edelman_Arias_Smith_1998}. Various works have appeared since that for classical and quantum control. For quantum control and quantum technologies, the first development of gradient and Hessian-based optimization theory on complex Stiefel manifolds was made in~\cite{Pechen_Prokhorenko_Wu_Rabitz_2008,Oza_Pechen_Dominy_Beltrani_Moore_Rabitz_2009}, where quantum control and optimization of open systems was reformulated as Riemannian optimization over complex Stiefel manifolds (strictly speaking, over some factors of complex Stiefel manifolds over some equivalence relation). In~\cite{Pechen_Prokhorenko_Wu_Rabitz_2008} (arXiv preprint of 2007) it was developed  for two-level quantum systems, and in~\cite{Oza_Pechen_Dominy_Beltrani_Moore_Rabitz_2009} for general $N$-level quantum  systems.  In these works:
\begin{itemize}
	\item quantum control and optimization of open systems was formulated as Riemannian optimization over (some factors of) complex Stiefel manifolds;
	\item gradient and Hessian for a general class of quantum control objectives were explicitly computed;
	\item all critical points of these objectives where found and characterized~\cite{Wu_Pechen_Rabitz_Hsieh_Tsou_2008};
	\item constrained optimization was studied.
\end{itemize}

First, we noted that using Kraus operator-sum representation for any CPTP map~(\ref{Kraus1}) for an $N$-level quantum system, one can construct the following $N^3 \times N$ matrix
\[
S=
\left(
\begin{array}{c}
	K_1 \\ K_2 \\ \dots \\ K_{N^2}
\end{array}
\right)
\]
The condition~(\ref{Kraus2}) implies that $S^\dagger S=\mathbb I_{N}$ and defines exactly that is known in geometry as complex Stiefel manifold ${\cal S} = V_N(\mathbb{C}^{N^3})$. Hence any Kraus operator-sum representation corresponds to some point of the complex Stiefel manifold.  As is known, Kraus OSR is non-unique. Taking this non-uniqueness into account leads to factoring of the Stiefel manifold over some equivalence relation~\cite{Oza_Pechen_Dominy_Beltrani_Moore_Rabitz_2009}. 

Then the objective function for maximization of average value of a quantum observable $\cal O$ has the form
\[
J_\O = \Tr \left[\sum_{i=1}^{N^2} \tilde{K_{i}}\sigma\tilde{K}_{i}^{\dagger}\O\right]=\Tr\left[S\rho S^{\dagger}({\mathbb I}_{N^2}\otimes{\O})\right]
\]
Many quantum control problems can be reduced to such control objectives. 

One of the several key results of the approach~\cite{Oza_Pechen_Dominy_Beltrani_Moore_Rabitz_2009}  are the explicit expressions for gradient, and even more, Hessian of the objective on the complex Stiefel manifold, the objects which are necessary for performing optimization. These expressions  are exact and analytic.

\begin{theorem} Gradient and Hessian of the objective $J_\O$ have the form
\begin{eqnarray}
	{\rm grad}\, J_\O(S) &=&(2{\mathbb I}_{N^3}-SS^{\dagger})({\mathbb I}_{N^2}\otimes\O) S \rho - S\rho S^{\dagger}({\mathbb I}_{N^2}\otimes\O) S. \nonumber\\
	{\rm Hess }\, J_\O(S)(\delta S) &=& 2({\mathbb I}_{N^2}\otimes\O) (\delta S)\rho - (\delta S)S^{\dagger} ({\mathbb I}_{N^2}\otimes\O) S\rho - (\delta S)\rho S^{\dagger}({\mathbb I}_{N^2}\otimes\O) S \nonumber \\
	&&- SS^{\dagger}({\mathbb I}_{N^2}\otimes\O) (\delta S)\rho + SS^{\dagger}(\delta S)S^{\dagger}({\mathbb I}_{N^2}\otimes\O) S\rho \nonumber \\
	&& - S\rho(\delta S)^{\dagger}({\mathbb I}_{N^2}\otimes\O) S
	+ ({\mathbb I}_{N^2}\otimes\O) S\rho(\delta S)^{\dagger}S. \nonumber
\end{eqnarray}
\end{theorem}

For optimization, it is important to know all critical points of the objective, in particular, does it has local but not global optima. In this regard, without explicit use of the Stiefel manifolds all critical points of such objectives where found and characterized, and it was shown that they can be only global maxima, global minima, and saddles~\cite{Wu_Pechen_Rabitz_Hsieh_Tsou_2008}. Constrained optimization was also studied~\cite{Oza_Pechen_Dominy_Beltrani_Moore_Rabitz_2009}.

\section{Reachable states for an open qubit}\label{secReachable}

The first key result to be ideally established for any given control system is to describe its controllability~\cite{Schirmer_Fu_Solomon_2001,Turinici_Rabitz_2001,Polack_Suchowski_Tannor_2009}, and, if the system is not controllable, its reachable sets, i.e., sets of states which are reachable from a given initial state in {\it any} time using {\it any} admissible (in our case coherent and incoherent) controls. In Ref.~\cite{Lokutsievskiy_Pechen_2021}, which also contains an overview of various results on controllability and analysis of reachable sets for quantum systems, this problem was analytically solved  for a qubit interacting with the environment and driven by a single coherent control with (and without) incoherent controls in the Bloch ball representation of quantum states using the technique of geometric control theory. 

The following master equation was considered for an open two-level quantum system (qubit) driven by coherent and incoherent controls with time-dependent decoherence rate determined by $\gamma(t) = \gamma [n(t)+1/2]$, where $n(t)$ is the spectral density of incoherent photons surrounding the qubit at the time~$t$:
\begin{equation}
	\frac{d\rho}{dt}  = - i [H_0 + V u(t), \rho] + \gamma \mathcal{L}_{n(t)}(\rho),\quad \rho(0) = \rho_0.
	\label{system_qubit}
\end{equation}
Here, $H_0 = \omega \begin{pmatrix}
	0 & 0 \\
	0 & 1
\end{pmatrix}$ is the free Hamiltonian, 
$V = \mu\sigma_x
= \mu\begin{pmatrix}
	0 & 1 \\
	1 & 0
\end{pmatrix}$ is the interaction Hamiltonian, $\omega$ is the transition frequency, $\mu>0$ is the dipole moment, $u(t)$ is coherent control (real-valued function), and $\gamma>0$ is the decoherence rate coefficient. The~dissipative superoperator is
\begin{multline*} 
	\mathcal{L}_{n(t)}(\rho_t) = n(t) \left(\sigma^+\rho_t\sigma^- + \sigma^-\rho_t\sigma^+ - \dfrac{1}{2}\{\sigma^-\sigma^+ + \sigma^+\sigma^- , \rho_t\}\right) \\ + \left(\sigma^+\rho_t\sigma^- -  \dfrac{1}{2}\{\sigma^-\sigma^+, \rho_t\}\right),
\end{multline*}
where $n(t) \geq 0$ is incoherent control (non-negative real-valued function), matrices $\sigma^\pm = \dfrac{1}{2} (\sigma_x \pm i\sigma_y)$, $\sigma^+ = \begin{pmatrix}
	0 & 1 \\
	0 & 0
\end{pmatrix}$, $\sigma^-= \begin{pmatrix}
	0 & 0 \\
	1 & 0
\end{pmatrix}$, and~$\sigma_x$, $\sigma_y$, $\sigma_z$ are the Pauli~matrices. 
Then the set of reachable states was analytically described. In particular, it was shown that most states can be reached exactly, while there exist a special set of unreachable states. Its maximal size was found to be $\delta \gamma/\omega$, where $\delta\approx 1$, and explicit description of this set was provided.  An exact description of the reachable sets, as well as its visualization in the Bloch ball, are provided in~\cite{Lokutsievskiy_Pechen_2021}.

\section{Incoherent GRAPE optimization and quantum control landscapes}
The reachable set gives an answer about {\it existence} of controls  steering initial state to a final state. For practical applications, after the set of reachable states is known, the next problem is how to find a control which steers the initial state to a given state from the reachable set. For this task, we developed incoherent version of the GRAPE (inGRAPE) approach (which was originally developed in the seminal work~\cite{Khaneja2005} for optimizing electromagnetic pulses for NMR) for the problem of coherent and incoherent control of open quantum systems~\cite{Petruhanov_Pechen_2023JPA}. For a single qubit under coherent and incoherent controls, an exact analytical expression for the dynamics and gradient for piece-wise constant controls was obtained using solutions of a third order equation via Cardano method. As the next stage, efficiency of the method was estimated for various tasks, as for example single-qubit~\cite{Petruhanov_Pechen_2023Photonics} and two-qubit~\cite{Pechen_etal} gate generation. Interesting is that for $T$ (or $\pi/8$) gate optimization of various objectives describing gate generation (e.g., based on the GRK approach developed by M.Y.~Goerz, D.M.~Reich, and C.P.~Koch~\cite{Goerz_NJP_2014_2021}) from randomly distributed initial conditions results in the distribution with two different minimal infidelity values and with two different groups of optimized controls. Whereas for Hadamard gate, C-NOT and C-Z gates such optimization results in smooth landscape with only one maximum~\cite{Pechen_etal}. The reason for such different behavior has to be studied.

\end{document}